\documentclass[
superscriptaddress,
 preprint,
 amsmath,amssymb,
 aps,
]{revtex4}

\usepackage{graphicx}
\usepackage{dcolumn}
\usepackage{bm}
\usepackage{braket,color}
\usepackage[colorlinks=true, urlcolor=blue, linkcolor=blue, citecolor=blue, bookmarks, pdftitle={article}]{hyperref}

\newcommand{\Tom}[1]{#1}

\begin{document}


\title{Exciton g-factors of van der Waals heterostructures from first principles calculations}

\author{Tomasz Wo{\'z}niak}
\email{Tomasz.wozniak@pwr.edu.pl}
\affiliation{Department of Semiconductor Materials Engineering, Wroc\l{}aw University of Science and Technology, 
50-370 Wroc\l{}aw, Poland}

\author{Paulo E.~Faria Junior}
\affiliation{Institute for Theoretical Physics, University of Regensburg, 93040 Regensburg, Germany}

\author{Gotthard Seifert}
\affiliation{Theoretical Chemistry, TU Dresden, 01062 Dresden, Germany}

\author{Andrey Chaves}
\affiliation{Departamento de Fisica, Universidade Federal do Cear\'a, 
60455-900 Fortaleza, Cear\'a, Brazil}
\affiliation{Department of Physics, University of Antwerp, Groenenborgerlaan 171, B-2020 Antwerpen, Belgium}

\author{Jens Kunstmann}
\email{jens.kunstmann@tu-dresden.de}
\homepage{http://www.j-kunstmann.de/}
\affiliation{Theoretical Chemistry, TU Dresden, 01062 Dresden, Germany}

\date{\today}

\begin{abstract}
External fields are a powerful tool to probe optical excitations in a material. The linear energy shift of an excitation in a magnetic field is quantified by its effective g-factor.
Here we show how exciton g-factors and their sign can be determined by converged first principles calculations. We apply the method to monolayer excitons in semiconducting transition metal dichalcogenides 
and to interlayer excitons in MoSe$_2$/WSe$_2$ heterobilayers and obtain good agreement with recent experimental data. 
The precision of our method allows to assign measured g-factors of optical peaks to specific transitions in the band structure and also to specific regions of the samples. This revealed the nature of various, previously measured interlayer exciton peaks.   
We further show that, due to specific optical selection rules, g-factors in van der Waals heterostructures are strongly spin- and stacking-dependent.
The calculation of orbital angular momenta requires the summation over hundreds of bands, indicating that for the considered two-dimensional materials the basis set size is a critical numerical issue. 
The presented approach can potentially be applied to a wide variety of semiconductors.
\end{abstract}

\maketitle

\section{Introduction}

Since the dawn of quantum mechanics the application of external magnetic fields has proven to be an  invaluable tool to probe the properties of matter. 
A good textbook example is the Zeeman effect in atoms, that describes the linear shift of an energy level $\varepsilon = g \mu_\mathrm{B} B$ in a homogeneous magnetic field $B$, where $g$ is the {Land\'e} g-factor and $\mu_\mathrm{B}$ is the Bohr magneton.
The theory of magnetic field shifts in semiconductors is closely related and was developed  by multiple authors before \cite{Roth1959,Hermann1977a,FariaJunior2019}, mostly within the context of k$\cdot$p perturbation theory or few-band tight-binding models. For conventional semiconductors, these models have proven to be useful and predictive but their applications to two-dimensional semiconductors based on transition metal dichalcogenides (TMD) has not led to satisfactory results yet \cite{Kormanyos2015,Wang2015f,Rybkovskiy2017}. {Early experimental studies of the magnetic field dependence of excitons, i.e.~optical excitations formed by bound electron-hole pairs, in monolayer MoSe$_2$ observed a Zeeman shift g $\approx$ 4, which has been attributed to the $d$-orbital character of the conduction and valence states involved in the excitonic transition \cite{Li2014h,Macneill2015}. However, subsequent studies in WSe$_2$ \cite{Srivastava2015b,Stier2018a,Chen2019} and WS$_2$ \cite{Zipfel2018a} where excitons exhibit the same orbital character, observed slightly larger values, which pointed to possible corrections due to the angular momentum texture of the conduction and valence bands. This picture became even more puzzling when g-factors of $\approx$ 9.5 were experimentally observed for dark exciton states in bilayer WSe$_2$ \cite{Lindlau2018}, and when inter-layer excitons in heterobilayers of TMD where demonstrated to have g-factors of $\approx 6.7$ and $\approx - 16$ \cite{Seyler2019}, which deviate even more from the value expected for ground state excitons in TMD. It is thus clear that a more rigorous theoretical model, which properly accounts for the angular momentum character of conduction and valence states in monolayer and bilayer materials, is required for an accurate description of the exciton Zeeman shifts in these materials. In this work, we address this problem and offer a practical solution that particularly works for excitonic states.} 

\begin{figure}[tb]
\includegraphics[width=\columnwidth]{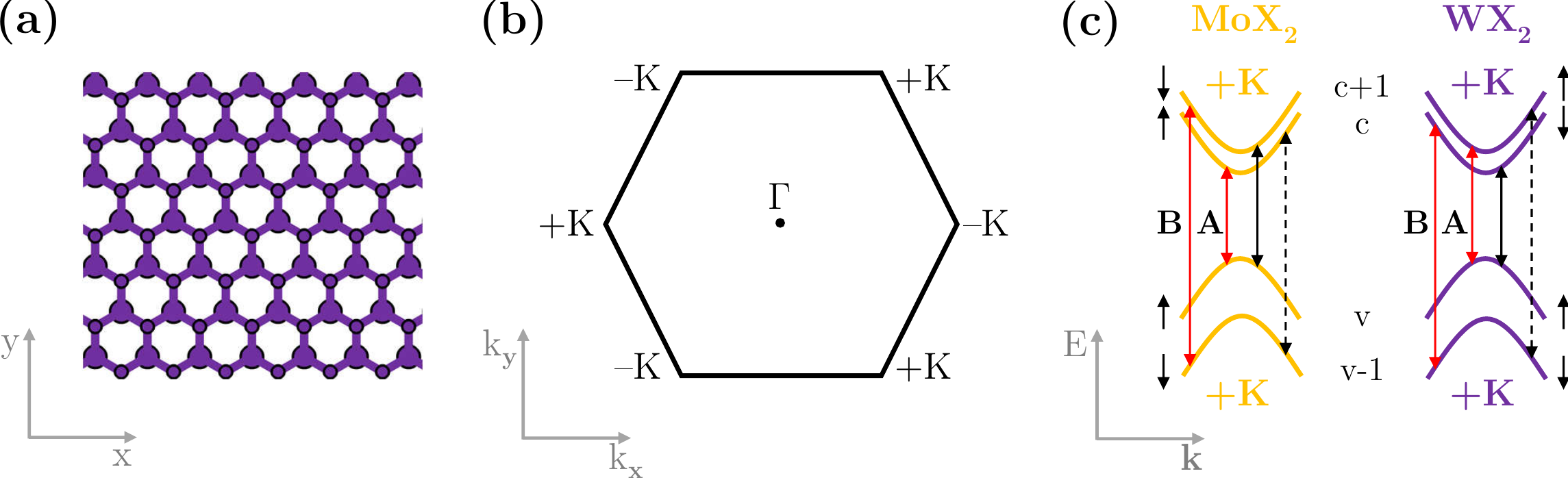}
\caption{\label{fig:1L} 
Properties of transition metal dichalcogenide monolayers MX$_2$. (a) Top view of the atomic structure, large and small balls represent M (metal) and X (chalcogen) atoms, respectively. 
(b) The Brillouin zone with the points $\Gamma$ at the center and K at the corners. The sign of the K points (valley index) alternates. 
(c) Schematic band structure at the +K point. Small arrows next to the colored bands indicate the spin orientation of the  conduction (c, c+1) and valence (v-1, v) bands. Double arrows indicate dipole-allowed optical  transitions, where the polarization $\sigma +$ is shown in red, $z$ in black and the dashed line represents a forbidden transition. In summary: the spin-conserving transitions at +K couple to $\sigma +$ polarized light, one of the spin-flip transitions is optically dark and the other one couples to $z$-polarized light.
}
\end{figure}

\begin{figure}[tb]
\includegraphics[width=\columnwidth]{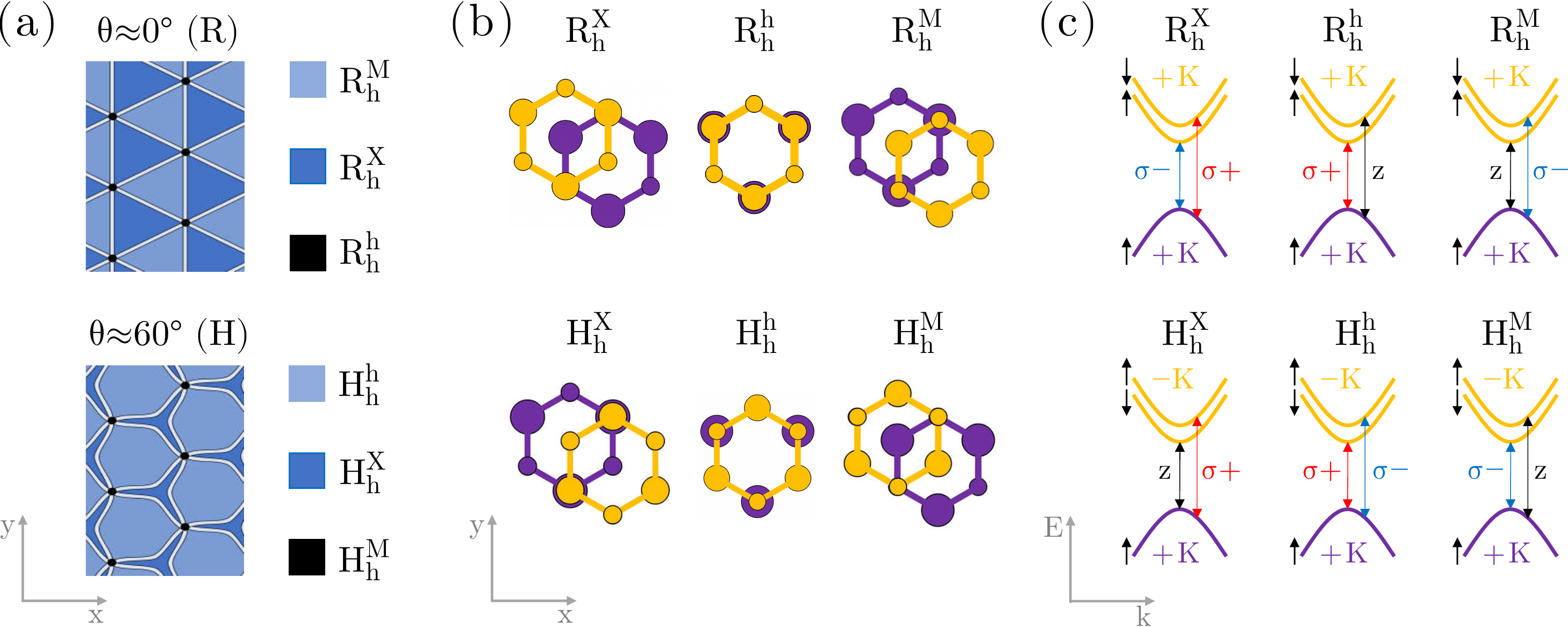}
\caption{\label{fig:HB} 
Properties of transition metal dichalcogenide heterobilayers for interlayer twist angles $\theta$ close to  0$^\circ$ (top line) and 60$^\circ$ (bottom line), as exemplified by MoSe$_2$/WSe$_2$.
(a) Scheme of the periodic atomic structure reconstruction, indicating strong deviations from ideal moir{\'e} patterns. The area of low-energy, high-symmetry stacking configurations (blue) is significantly enlarged and 0$^\circ$ and 60$^\circ$ have different reconstructions.
(b) The geometry of high-symmetry stacking configurations, where purple corresponds to WSe$_2$ and orange to MoSe$_2$ layers. Metal atoms are depicted by bigger circles and chalcogenes by smaller ones. 
(c) Schematic band structures of the stacking configurations  at the +K point of the heterobilayer Brillouin zone. The color code indicates that MoSe$_2$ is the electron layer and WSe$_2$ is the hole layer.
Small arrows to the left of the bands indicate the spin-orientation. Double arrows indicate dipole-allowed optical  transitions (selection rules), where  $\sigma +$, $\sigma -$ and $z$ are the corresponding polarizations. 
The selection rules are strongly spin- and stacking-dependent where, contrary to monolayers (see Fig.~\ref{fig:1L}), spin-flip transitions can couple to $\sigma +$ or $\sigma -$ polarized light.
}
\end{figure}

To test and apply the method we consider monolayers (see Fig.~\ref{fig:1L}) and heterobilayers (see Fig.~\ref{fig:HB}) of TMD.
They are particularly suited to our method because (i) their optical properties are dominated by excitons  and (ii) related phenomena such as exciton complexes, Rydberg series, Zeeman shifts and more were recently studied in great detail  \cite{Koperski2017a,Wang2018}.
A van der Waals heterostructure is formed by vertically stacking two-dimensional crystals via deposition or mechanical exfoliation. Today it is possible to fabricate heterostructures with arbitrary material sequence and relative lattice orientation (twist angle  $\theta$) \cite{Geim2013}. The interlayer interactions are weak and therefore many monolayer properties are preserved in heterostructures. TMD heterobilayers (HB) usually have a staggered (type-II) band alignment and free electrons and holes accumulate in different layers which leads to the formation of long-lived, charge-separated, spatially-indirect interlayer excitons  
\cite{Fang2014d,Rivera2014,Rivera2016}. 

A mismatch of the in-plane lattice constants or a sufficiently large twist angle  between individual layers leads to the formation of a moir{\'e} pattern where the lattice registry and the band gap continuously vary in space. This gap variation can act as an additional confining potential for interlayer excitons \cite{Zhang2017f,Nayak2017,Tran2019}. 
It was recently shown that in MoSe$_2$/WSe$_2$ HB and MoS$_2$ bilayers with $\theta$ close to 0$^\circ$ (R) or 60$^\circ$ (H) structural deformations lead to strong deviations from the ideal  moir{\'e} pattern and  the areas of high-symmetry stacking configurations with the lowest total energies 
are significantly enlarged \cite{Rosenberger2019,Weston2019}. The period of these  deformations is equal to the moir{\'e} wave length. 
For R systems the sample area is mostly covered by equal proportions of R$_h^X$ (AB) and R$_h^M$ (BA) stackings, while in H systems H$^h_h$ (ABBA) covers most of the sample \footnote{Similar to bulk TMD, the energetically favorable stackings are those where X and M atoms are vertically aligned.
}. This is illustrated in Fig.~\ref{fig:HB}(a,b) where for the labeling of the stacking configurations \footnote{The labels are R$_m^n$ and H$_m^n$. 
R and H correspond to 0$^\circ$  and 60$^\circ$ twist angle, respectively. The subscript $m$ refers to the hole layer  and the superscript $n$ to the electron layer. 
For $m$ and $n$ three specific positions of a monolayer lattice are considered: $h$ - hollow center of a hexagon, $X$ - chalcogene atom site and $M$ - metal atom site. So H$_h^M$ is the 60$^\circ$ stacking configuration where the metal site of the electron layer (MoSe$_2$) is over the hole site of the hole layer (WSe$_2$).
} 
we follow the notation of Yu \textit{et al.}~\cite{Yu2017c,Yu2018}.

In TMD monolayers, the fundamental band gap is direct and located at the  corners of the hexagonal Brillouin zone at the $\pm$K points (see Fig.~\ref{fig:1L}(b)). There are two symmetry inequivalent $\pm$K valleys, that are connected by time-reversal symmetry, and the sign is  called the valley index. Spin-orbit interactions split the band edge states into spin-polarized bands as indicated in Fig.~\ref{fig:1L}(c). The magnitude of the splitting is several hundred meV in the valence band and only a few meV in the conduction band. 
Due to mirror symmetry in monolayers, the projection of the spin onto the quantization axis perpendicular to the layer is preserved and $m_z = 1/2$ is a good quantum number. \Tom{However, the spin orbit coupling can lead to a reduction of $m_z$, while preserving $m_x = m_y = 0$, as shown in Ref. \cite{Kurpas2019} for 2D hexagonal crystals. Nevertheless, in most cases, taking $m_z = 1/2$ was demonstrated to be a reasonable approximation \cite{Xiao2012}.}
In  molybdenum-based monolayers the spin orientation of the valence and conduction bands is the same, while in tungsten-based systems the spin orientation is opposite \cite{Kormanyos2014}.
At the $\pm$K valleys optical transitions couple to light of specific
circular ($\sigma \pm$) or linear ($z$) polarization, as indicated by vertical double arrows in Fig.~\ref{fig:1L}(c).
The allowed transitions are determined by dipole selection rules
\begin{eqnarray}
| \mathbf{e_+} \cdot \bm{\pi}_{cv\mathbf{k}} |^2 > 0 &\longleftrightarrow \sigma+, \nonumber\\
| \mathbf{e_-} \cdot \bm{\pi}_{cv\mathbf{k}} |^2 > 0 &\longleftrightarrow \sigma-, \label{eqn:sel_rules}\\
| \mathbf{z} \cdot \bm{\pi}_{cv\mathbf{k}} |^2 > 0 &\longleftrightarrow z, \nonumber
\end{eqnarray} 
where $\mathbf{e_\pm} = (1,\pm i,0)/\sqrt{2} $, $\mathbf{z}=(0,0,1)$ and $\bm{\pi}_{cv\mathbf{k}} = (\pi^x_{cv\mathbf{k}}, \pi^y_{cv\mathbf{k}}, \pi^z_{cv\mathbf{k}})$ are momentum (or optical) matrix elements for transitions between the valence and conduction band and $v$, $c$ are the corresponding band indices.
The left-hand side of Eq.~(\ref{eqn:sel_rules}) is directly proportional to the oscillator strength of a transition and therefore we will refer to it as "intensity".
The selection rules differ in monolayers and HB, where they are also stacking-dependent \cite{Yu2018}. 
In Fig.~\ref{fig:1L}(c) it is discernible that in monolayers the spin-conserving transition (giving rise to spin-singlet excitons) couples to $\sigma +$ light at the +K valley (and to $\sigma -$ at --K) and one spin-flip transition (leading to spin-triplet excitons) couples to $z$-polarized light and the other one is forbidden/dark. In stark contrast are the selection rules of MoSe$_2$/WSe$_2$ HB, that are shown in Fig.~\ref{fig:HB}(c). There, depending on the stacking configuration, spin-conserving and spin-flip transitions couple to entirely different polarizations, e.g., for the spin-conserving transition in a $R_h^h$ HB we have ($\sigma \pm \leftrightarrow \pm$K),
while in a $R_h^X$ HB we have ($\sigma \pm \leftrightarrow \mp$K).

In this paper we demonstrate how the theory of magnetic field-induced energy shifts in semiconductors can be realized with state of the art density functional theory calculations.
We test the method by calculating g-factors of excitons in MoS$_2$, MoSe$_2$, MoTe$_2$, WS$_2$, WSe$_2$ monolayers and obtain excellent agreement with available experimental data.
Then, we consider interlayer excitons in 
MoSe$_2$/WSe$_2$ HB (which might serve as model for arbitrary TMD-based HB) and show that the approach can explain recent magnetooptical measurements on HB, where unusual signs and values of excitonic g-factors were reported 
\cite{Nagler2017,Ciarrocchi2019,Seyler2019,Wang2019}.
We further demonstrate how stacking-dependent selection rules lead to stacking dependent exciton g-factors.

\section{Theory of magnetic field shifts in semiconductors}
\subsection{Effective g-factor of a Bloch state}
The basic theory {of the magnetic field dependence of Bloch states} has been developed before by multiple authors and is usually applied in models \cite{Roth1959,Hermann1977a,Kormanyos2015,Wang2015f,Rybkovskiy2017,FariaJunior2019,Junior2019}. Here we reformulate it in a way suitable for general electronic structure calculations.
The starting point is a non-relativistic band structure Hamiltonian $H^0$ and its corresponding band energies $\varepsilon^0_{n \mathbf{k}}$ and Bloch states $|n \mathbf{k} \rangle$ (i.e.~Bloch phase times lattice-periodic function)
\begin{align}
    H^0 &= \frac{\mathbf{p}^2}{2m_0} + V \label{eqn:H0}\\
    H^0 &\ |n \mathbf{k} \rangle = \varepsilon_{n \mathbf{k}}^0 |n \mathbf{k} \rangle \\
     \mathbf{1} &= \sum_n  |n \mathbf{k} \rangle \langle n \mathbf{k} |, \label{eqn:1}
\end{align}
where $\mathbf{p}$ is the momentum operator, $m_0$ is the rest mass of the electron, $V$ is the effective potential and  $n$ and $\mathbf{k}$ are the band index and the wave number, respectively. 
The last line emphasizes that the set of Bloch states forms a complete basis. These states are obtained from electronic structure calculations and are supposed to be known.
The coupling of these states to an external magnetic field is described by adding the spin Zeeman term to $H^0$ and by replacing the momentum operator $\mathbf{p}$ by $\mathbf{p} - q \mathbf{A}$ (minimal coupling), where $\mathbf{A}$ is the vector potential, $q=-|e_0|$ the charge of the electron and $e_0$ is the elementary charge. For a uniform external magnetic field $\mathbf{B}$ it is convenient to choose  $\mathbf{A} = (\mathbf{B} \times \mathbf{r})/2$, which satisfies the Coulomb gauge $\bm{\nabla} \cdot \mathbf{A} = 0$, where $\mathbf{r}$ is the position operator.  This leads to the Pauli equation 
\begin{align}
H(\mathbf{B}) &= H^0 + H^\mathrm{L}(\mathbf{B}) + H^\mathrm{Q}(\mathbf{B}) \nonumber\\
  &= H^0 + \mu_\mathrm{B} \mathbf{B} \cdot \left( \mathbf{L} + \frac{g_0}{2}  \bm{\Sigma} \right) + \frac{e_0^2}{8m_0} \left( \mathbf{B} \times \mathbf{r} \right)^2 \label{eqn:Pauli},
\end{align}
where $\mu_\mathrm{B} = \hbar e_0/2m_0$ is the Bohr magneton, $ \mathbf{L} = (\mathbf{r} \times \mathbf{p})/\hbar$ is the (dimensionless) angular momentum operator,  $\bm{\Sigma} = (\Sigma^x, \Sigma^y, \Sigma^z)$ is the vector of Pauli matrices, and $g_0$ is the g-factor of the free electron.
Above, we separate Eq.~(\ref{eqn:Pauli}) into  $H^\mathrm{L}(\mathbf{B})$ and $H^\mathrm{Q}(\mathbf{B})$ that represent the part of $H(\mathbf{B})$ that linearly and quadratically depend on $\mathbf{B}$, respectively.

Let us now consider that for a band edge state of a semiconductor the eigenvalues $\varepsilon_{n \mathbf{k}}$ are of the order of 1 eV. It is further experimentally known that for a field of $B \approx 10$ T the energy shifts of the band energies are of the order of 1 meV. Thus $H^\mathrm{L}(\mathbf{B})$ and $H^\mathrm{Q}(\mathbf{B})$ are weak perturbations of $H^0$ and the magnetic field shift of the band energies can be evaluated with first order perturbation theory. This gives
\begin{align*}
\varepsilon_{n \mathbf{k}}(\mathbf{B}) = \varepsilon_{n \mathbf{k}}^0 + \langle n \mathbf{k} |H^\mathrm{L}(\mathbf{B}) + H^\mathrm{Q}(\mathbf{B}) |n \mathbf{k} \rangle.
\end{align*}
Now choosing $\mathbf{B} = (0,0,B)$ parallel to the Cartesian $z$ direction and  $g_0/2 \approx 1$ we get
\begin{align}
\varepsilon_{n \mathbf{k}}(B) = \varepsilon_{n \mathbf{k}}^0 + \mu_\mathrm{B} B \left( L_{n \mathbf{k}} + \Sigma_{n \mathbf{k}} \right) + H_{n \mathbf{k}}^\mathrm{Q},
\label{eqn:enk}
\end{align}
with the matrix elements $L_{n \mathbf{k}}= \langle n \mathbf{k} |L^z |n \mathbf{k} \rangle$, $\Sigma_{n \mathbf{k}}= \langle n \mathbf{k} |\Sigma^z |n \mathbf{k} \rangle$ and $H_{n \mathbf{k}}^\mathrm{Q}= e_0^2 B^2/ 8m_0 \\ \langle n \mathbf{k} | {( r^x )}^2 + {( r^y )}^2 |n \mathbf{k} \rangle$. The effective g-factor of the Bloch state  $|n \mathbf{k} \rangle$ is thus
\begin{align}
g_{n \mathbf{k}} = L_{n \mathbf{k}} + \Sigma_{n \mathbf{k}}.
\label{eqn:gnk}
\end{align}

The orbital angular momentum matrix elements are evaluated as
\begin{align}
L_{n \mathbf{k}} &= \frac{1}{\hbar}  \langle n \mathbf{k} | r^x p^y - r^y p^x |n \mathbf{k} \rangle \nonumber \\
&= \frac{1}{\hbar}  \sum_{m=1}^N  r^x_{nm\mathbf{k}} p^y_{mn\mathbf{k}} - r^y_{nm\mathbf{k}} p^x_{mn\mathbf{k}} \nonumber\\
&= \frac{1}{i m_0} \sum_{m=1, m \neq n}^N \frac{p^x_{nm\mathbf{k}} p^y_{mn\mathbf{k}} - p^y_{nm\mathbf{k}} p^x_{mn\mathbf{k}}}{\varepsilon_{n \mathbf{k}} - \varepsilon_{m \mathbf{k}}} \label{eqn:L},
\end{align}
with the matrix elements $r^{\alpha}_{nm\mathbf{k}} = \langle n \mathbf{k} | r^{\alpha} |m \mathbf{k} \rangle$, $p^{\alpha}_{nm \mathbf{k}} = \langle n \mathbf{k} | p^{\alpha} |m \mathbf{k} \rangle$, where $\alpha=x,y,z$ represents Cartesian components.
The step from the first to the second line involves the insertion of the identity operator (\ref{eqn:1}), $r^x p^y = r^x \mathbf{1} p^y$,  where the basis contains $N$ states. Mind that the identity is only fulfilled if $N$ is sufficiently large (see discussion below). 
The second line involves the matrix elements of the position operator, that are non-trivial to evaluate in periodic systems  \cite{Thonhauser2011,Xiao2010}.
This problem is circumvented by using the commutator relation 
$[H^0, \mathbf{r}] = \frac{ \hbar}{i m_0} \mathbf{p}$, 
that can explicitly be shown to hold. Taking its matrix elements one finds
$r^{\alpha}_{nm\mathbf{k}} = \frac{\hbar}{i m_0} \frac{p^{\alpha}_{nm\mathbf{k}}}{\varepsilon_{n \mathbf{k}} - \varepsilon_{m \mathbf{k}}}$, 
$\varepsilon_{n \mathbf{k}} \neq \varepsilon_{m \mathbf{k}}$ and obtains Eq. (\ref{eqn:L}).
%
The band energies $\varepsilon_{n \mathbf{k}}$ and the matrix elements $\Sigma_{n \mathbf{k}}$ and $p^{\alpha}_{nm\mathbf{k}}$ can be obtained from electronic structure calculations and hence allow to calculate the effective g-factor of a Bloch state  $g_{n \mathbf{k}}$ (Eq. (\ref{eqn:gnk})). 
An alternative derivation of Eqs. (\ref{eqn:enk})-(\ref{eqn:L}) can be obtained with the semiclassical theory of Bloch electron dynamics in the presence of external fields, where the band energies are corrected by the magnetic moments as in Eq. (\ref{eqn:enk}) and the Berry curvature appears as a correction to the group velocity in the equations of motion \cite{Chang1995, Xiao2010}. In this theory the orbital moment can be seen as a self-rotation of a Bloch wave packet around its center of mass.

Equation (\ref{eqn:L}) {can be applied not only to Bloch states of crystals, but also to atoms or molecules}. For the hydrogen atom it can be shown that for a sufficiently large number of states $N$, included in the summation, this expression converges to the well known analytical result 
$L_{n'l'm'} = \langle n'l'm' | L_z | n'l'm' \rangle = m'$ \footnote{M.~M.~Glazov (Ioffe Institute, St.Petersburg, Russia), private communication}. However, the convergence is slow. 
In  the literature on TMD $L_{n \mathbf{k}}$ is sometimes divided into a contribution coming from the atomic orbital (ao) and one from the lattice (l) (or valley) 
$L_{n \mathbf{k}} = L_{n \mathbf{k}}^\mathrm{ao} + L_{n \mathbf{k}}^\mathrm{l}$ 
and the two contributions are separately discussed \cite{Srivastava2015b,Nagler2017,Chen2019,Seyler2019}. However, this division is only of qualitative nature, as the projection of a Bloch state $|n \mathbf{k} \rangle$ onto atomic-like orbitals is non-unique and leads to contributions from multiple atomic-like orbitals. 

\subsubsection{Relativistic effects}
Above, we outlined the non-relativistic theory that is satisfactory for 
light elements, but for systems with heavier atoms (such as Mo and W) 
relativistic effects cannot be neglected.  In this paper we are mostly concerned with electronic structure calculations based on density functional theory (DFT).
Relativistic effects and external magnetic fields can be introduced into DFT  via  current density functional theory \cite{Rajagopal1973,Eschrig1985}.
However, for valence states it is sufficient to consider 
a 2-spinor formulation for an approximate relativistic Hamiltonian $H^\mathrm{0,rel} = H^0(\mathbf{p}^2) + H^\mathrm{SOC}(\mathbf{p}) + H^\mathrm{MV}(\mathbf{p}^4) + H^\mathrm{D} + mc^2$, where $H^0$ is Hamiltonian (\ref{eqn:H0}) and the other terms represent the spin-orbit coupling (SOC), the mass-velocity relation, the Darwin shift and the electron rest mass, respectively \cite{SchwablQM}. 
Neglecting the spin-orbit term leads to a scalar-relativistic approach, that is often used in solid state codes \cite{Koelling1977}. 

In g-factor calculations including relativistic effects $H^0$ in Eq. (\ref{eqn:H0}) is replaced by $H^\mathrm{0,rel}$ which defines the set of unperturbed Bloch states.
Then the coupling of $H^\mathrm{0,rel}$ to the magnetic field is again realized by adding the spin Zeeman term and replacing $\mathbf{p}$ by $\mathbf{p} - q \mathbf{A}$ in the  parts that explicitly depend on $\mathbf{p}$. For $H^0(\mathbf{p}^2)$ this procedure  leads to Eq. (\ref{eqn:Pauli}). In TMD systems the coupling of $H^\mathrm{MV}(\mathbf{p}^4)$ leads to marginal corrections that are neglected here. This leaves $H^\mathrm{SOC}(\mathbf{p})$, which gives an additional linear contribution that is taken into account by replacing the momentum operator $\mathbf{p}$ in $H^\mathrm{L}(\mathbf{B})$  by \cite{Roth1959}
\begin{align}
\bm{\pi} = \mathbf{p} + \frac{\hbar}{4 m_0 c^2} \bm{\Sigma} \times \bm{\nabla} V.
\label{eqn:pi}
\end{align}
Specifically, ${p}^{\alpha}_{nm \mathbf{k}}$ needs to be replaced by ${\pi}^{\alpha}_{nm \mathbf{k}} = \langle n \mathbf{k} | \pi^{\alpha} |m \mathbf{k} \rangle$ in Eq. (\ref{eqn:L}).
Mind that this replacement also affects the optical selection rules (see Eq. (\ref{eqn:sel_rules})), where SOC enables spin-flip transitions.

\subsection{Effective g-factor of excitons}
Excitons are bound states formed by electron and holes from the conduction (c) and valence (v) band edges, respectively. Using expression (\ref{eqn:enk}) we define the momentum-direct exciton energy as
\begin{align}
E_\mathbf{k}(B) &= \varepsilon_{c \mathbf{k}}(B) - \varepsilon_{v \mathbf{k}}(B) - E_\mathbf{k}^\mathrm{Binding} \nonumber\\
&= E_\mathbf{k}^0 + E_\mathbf{k}^\mathrm{L}(B) + E_\mathbf{k}^\mathrm{Q}(B),
\label{eqn:Eex}
\end{align}
where $E_\mathbf{k}^\mathrm{Binding}$ is the exciton binding energy (that varies throughout the Brillouin zone), $E_\mathbf{k}^0 = \varepsilon_{c \mathbf{k}}^0 - \varepsilon_{v \mathbf{k}}^0 - E_\mathbf{k}^\mathrm{Binding}$ is the zero-field exciton energy, $E_\mathbf{k}^\mathrm{Q}(B) = H_{c \mathbf{k}}^\mathrm{Q} - H_{v \mathbf{k}}^\mathrm{Q}$ is the quadratic shift. The linear shift is 
\begin{align}
E_\mathbf{k}^\mathrm{L}(B) = (g_{c \mathbf{k}} - g_{v \mathbf{k}}) \mu_\mathrm{B} B  = g_{\mathbf{k}} \mu_\mathrm{B} B   \label{eqn:gex}
\end{align}
and $g_{\mathbf{k}}$ is the  \textit{intra-valley} g-factor of an exciton at  $\mathbf{k}$.

It is also possible to consider momentum-indirect excitons, where electron and hole originate from Bloch states with different crystal momentum $\mathbf{k}$ \cite{Kunstmann2018}.

\section{Numerical methods}
The electronic structure calculations were performed with density functional theory (DFT) using the Vienna Ab Initio Simulation Package (VASP) \cite{VASP} version 5.4.4, Perdew-Burke-Ernzerhof (PBE) \cite{PBE} exchange-correlation functional and the Projector Augmented Wave method \cite{PAW} with  potentials of version 54. For testing purposes, we also used the local density approximation (LDA). An energy cutoff of 300 eV and a $6\times6\times1$ k-mesh were chosen after careful convergence tests. The k-space integration was carried out with a Gaussian smearing method using an energy width of 0.05 eV for all calculations. All unit cells were built with at least 15 {\AA} separation between replicates in the perpendicular direction to achieve negligible interaction. 
Dispersion interactions corrections were of Tkachenko-Scheffler (TS) type \cite{TS}. Atomic positions and lattice constants were optimized with $10^{-3}$ eV/\AA ~and 0.1 kbar precision. The optimized values are given in footnote 
\footnote{MoS${}_2$: 3.158\AA, MoSe${}_2$: 3.295\AA, MoTe${}_2$: 3.521\AA, WS${}_2$: 3.165\AA, WSe2${}_2$: 3.299\AA}.
A comparative calculation for WS$_2$ was performed with the all-electron, full-potential linearised augmented plane wave (LAPW) method as implemented in the ELK package, using default parameters \cite{ELK}. 
The momentum matrix elements $\pi^{\alpha}_{nm \mathbf{k}}$ in VASP 
were obtained from the wave function derivatives that 
are calculated within density functional perturbation theory \cite{Gajdos2006}, in ELK they were calculated according to Eq.~\ref{eqn:pi}.

\section{Results and discussion}

\subsection{Transition metal dichalcogenide systems and the impact of optical selection rules on g-factors}
\label{sec:TMD}
In TMD monolayers and heterostructures the band edge states are mostly at $\mathbf{k} = \pm$K, which is what we will focus on in this article.
Due to time-reversal symmetry $\Sigma_{n,+K} = -\Sigma_{n,-K}$ and  $L_{n,+K} = -L_{n,-K}$. 
Spin-orbit interactions split the band edge states of monolayers into spin-polarized bands (See Fig.~\ref{fig:1L}(c)) and $\Sigma_{v,\pm K} = \pm 1$ \Tom{is commonly assumed} \cite{Wang2018}. We use this specific property to \textit{define} the valley index; so the valley where the valence band maximum is spin-up is +K. \Tom{In fact, ab initio calculations of monolayer TMD show that $|\Sigma_{n,\pm K}| < 1$  at the band edge ($n=$v, v-1, c, c+1). However the effect is so small that it has a negligible influence on the g-factor \footnote{\Tom{
For TMD monolayers the spin contribution to the inter-valley g-factors $g_\mathrm{A}$ and $g_\mathrm{B}$ is: 0.004 and 0.016 for MoS$_2$; 0.008 and 0.020 for MoSe$_2$; 0.044 and 0.148 for WS$_2$; 0.056 and 0.188 for WSe$_2$, respectively. The values were calculated using results from Ref.~\cite{Zollner2019}}}. 
In TMD HB such calculations also show highly spin polarized band edge states at the K points \cite{Bussolotti2018}. Therefore taking $\Sigma_{n,\pm K} = \pm 1$ for those states is indeed a reasonable approximation.} 
For the Bloch state and exciton g-factors the above symmetry properties imply $g_{n,+K} = -g_{n,-K}$ and $g_{+K} = -g_{-K}$, respectively.

The valley-dependent selection rules, as discussed in the introduction and visualized in Figs.~\ref{fig:1L} and \ref{fig:HB}, are employed to experimentally determine the excitonic g-factors, where it is common to use
\begin{align}
E_{\sigma +}(B) - E_{\sigma -}(B) = g \mu_\mathrm{B} B
\label{eqn:g_exp}
\end{align}
to extract the linear magnetic shift and to define the \textit{inter-valley} g-factor $g$.
Using Eqs.~(\ref{eqn:Eex}) and (\ref{eqn:gex}) it follows for the lowest energy transition in MoS$_2$ monolayers (A exciton)
$g^\mathrm{1L}_\mathrm{A} = g_{\sigma +} - g_{\sigma -} = g_\mathrm{+K} - g_\mathrm{-K} = 2 g_\mathrm{+K}$. 
In a $R_h^h$ HB the selection rules are the same and we obtain the same result $g^{R_h^h} = 2 g_\mathrm{+K}$. 
But a $R_h^X$ HB has different selection rules and therefore $g^{R_h^X} = g_{\sigma +} - g_{\sigma -} = g_\mathrm{-K} - g_\mathrm{+K} = 2 g_\mathrm{-K}$. 
This demonstrates that in HB the inter-valley g-factors, as defined by (\ref{eqn:g_exp}), depend on the stacking configuration, which will further be discussed below.


\subsection{Exciton g-factors of monolayers}

\begin{figure}[htb]
\includegraphics[width=\columnwidth]{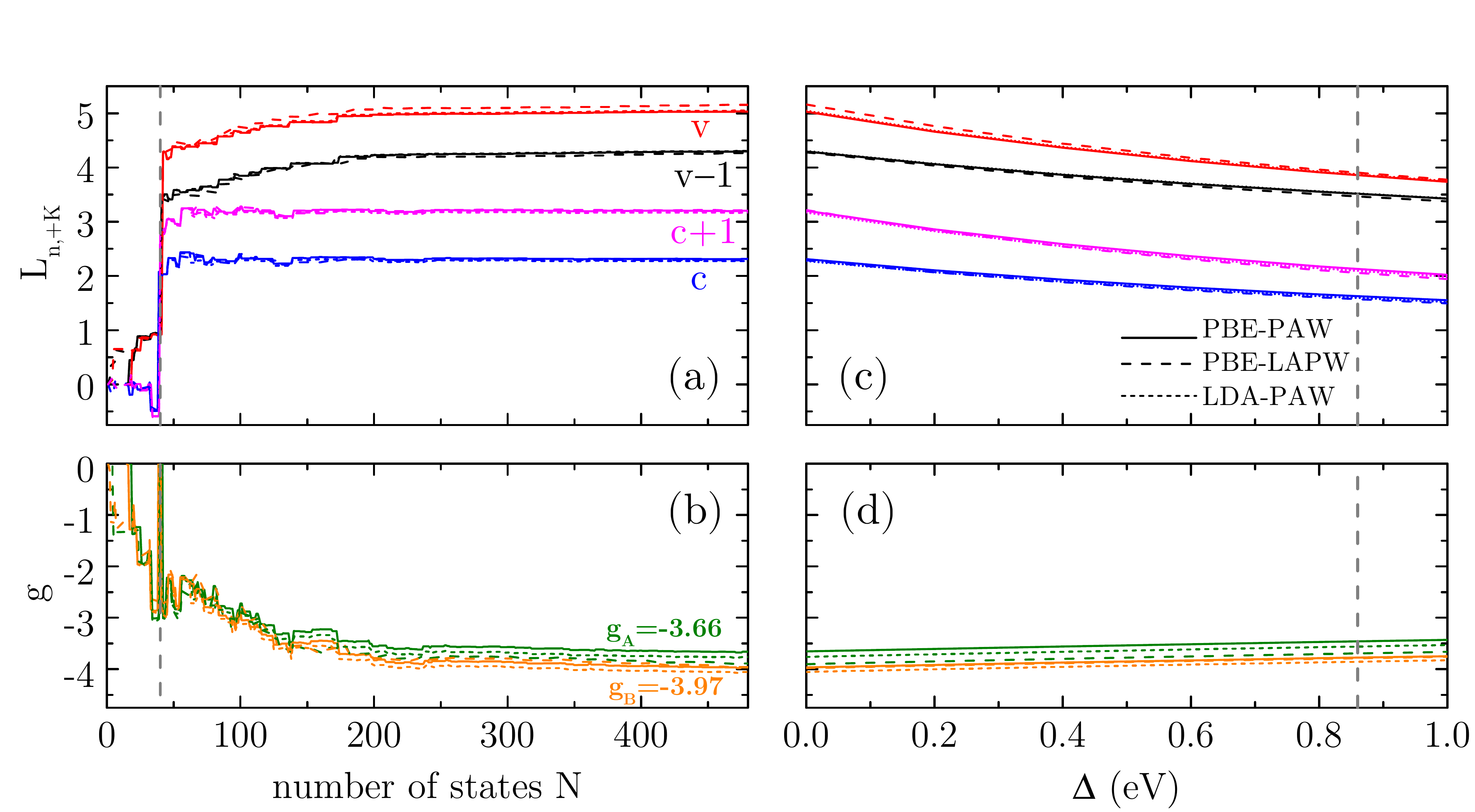}
\caption{\label{fig:conv} 
Impact of basis set size $N$, band gap correction $\Delta$, exchange-correlation functional (PBE, LDA) and electronic-structure method (PAW, LAPW) on orbital angular momenta and exciton g-factors in WS$_2$ monolayer.
(a) Convergence of the (dimensionless) orbital angular momenta $L_{n,+\mathrm{K}}$ of the two highest valence band states ($n=v,v-1$) and two lowest conduction band states ($n=c,c+1$) at the +K point with respect to the number of bands $N$ included in the calculation (Eq.~(\ref{eqn:L})).
(b) Convergence of the inter-valley g-factors of A and B excitons $g^\mathrm{1L}_\mathrm{A} = 2 (L_{c+1,+K} - L_{v,+K})$ and $g^\mathrm{1L}_\mathrm{B} = 2 (L_{c,+K} - L_{v-1,+K})$. 
$N=1$ is the lowest-energy state of the valence shell, the valence band maximum is indicated by a dashed vertical line. 
A large number of bands ($N \ge 300$) is required to converge the g-factors to a precision of 0.1.
(c) Impact of the band gap correction $\Delta$ on the orbital momenta $L_{n,+\mathrm{K}}$ and (d) the g-factor. The dashed vertical line indicates the G$_0$W$_0$ quasiparticle band gap. While the $L_{n,+\mathrm{K}}$ depend on $\Delta$, the exciton g-factors are almost insensitive to it.  
}
\end{figure}

\begin{table*}[htb]
\caption{\label{tab:1Ls}
Calculated g-factors of A and B excitons $g^\mathrm{1L}_\mathrm{A}$, $g^\mathrm{1L}_\mathrm{B}$ (Eq.~(\ref{eqn:g_exp})) in transition metal dichalcogenide monolayers and comparison with experimental literature values.
Despite the large spread of the experimental values the calculated results are in good agreement.
Also given are the related orbital angular momenta 
$L_n = L_{n,+\mathrm{K}}$ of the two lowest conduction band states ($n=c,c+1$) and highest valence band states ($n=v,v-1$) at the +K point
and the intensities $\frac{\hbar}{m_0} | \mathbf{e_+} \cdot \bm{\pi} |^2$ of the related circularly polarized transition in (eV$\cdot${\AA})$^2$ for the A exciton. \Tom{Using $|\Sigma_{n,+\mathrm{K}}| = 1$ leads to no spin contribution to the g-factor.}
All results are obtained with the PBE-PAW method.
}
\begin{ruledtabular}
\begin{tabular}{llllll}
 & MoS${}_2$ & MoSe${}_2$ & MoTe${}_2$ & WS${}_2$ & WSe${}_2$ \\
\hline
$g^\mathrm{1L}_\mathrm{A}$          & -3.68 & -3.82 & -3.96 & -3.66 & -3.80 \\
$g^\mathrm{1L}_\mathrm{A}$ (exp.)    & -1.7\footnotemark[1], -1.8\footnotemark[20], -2.9\footnotemark[2], & -3.8\footnotemark[5]${}^,$\footnotemark[6], -4.0\footnotemark[20], & -4.3\footnotemark[9], -4.7\footnotemark[9], & -3.7\footnotemark[16], -3.94\footnotemark[3], & -1.57\footnotemark[12], -2.86\footnotemark[12], -3.2\footnotemark[17], \\
                  &  -3.0\footnotemark[2], -3.6\footnotemark[2], -3.8\footnotemark[2], &  -4.1\footnotemark[7], -4.2\footnotemark[8], & -4.8\footnotemark[9] & -4.0\footnotemark[2]${}^,$\footnotemark[16], -4.25\footnotemark[10], & -3.7\footnotemark[6], -3.8\footnotemark[8], -4.1\footnotemark[22] \\
                  &  -4.0\footnotemark[3], -4.2\footnotemark[16], -4.6\footnotemark[4] & -4.3\footnotemark[2], -4.4\footnotemark[4] & & -4.35\footnotemark[11] & -4.25\footnotemark[19], -4.3\footnotemark[13], -4.37\footnotemark[14], \\
&&&&& -4.38\footnotemark[15] \\
$g^\mathrm{1L}_\mathrm{B}$ & -3.70 & -3.88 & -4.02 & -3.96 & -4.26 \\
$g^\mathrm{1L}_\mathrm{B}$ (exp.) & -4.3\footnotemark[4], -4.65\footnotemark[3] & -4.2\footnotemark[16] & -3.8\footnotemark[8] & -3.99\footnotemark[3], -4.9\footnotemark[16] & -3.9\footnotemark[16] \\
$L_{c} / L_{c+1}$  & 2.09/1.87 & 1.78/1.51 & 1.58/1.21 & 2.31/3.20 & 1.87/2.91 \\
$L_{v-1} / L_{v}$     & 3.72/3.93 & 3.45/3.69 & 3.22/3.56 & 4.29/5.03 & 4.00/4.81 \\
intensity (A) & 28.6 & 21.2 & 13.9 & 42.9 & 33.1 \\
\end{tabular}
\end{ruledtabular}
\footnotetext[1]{Ref. \cite{Cadiz2017}}
\footnotetext[2]{Ref. \cite{Goryca2019a}}
\footnotetext[3]{Ref. \cite{Stier2016a}}
\footnotetext[4]{Ref. \cite{Mitioglu2016}}
\footnotetext[5]{Ref. \cite{Macneill2015}}
\footnotetext[6]{Ref. \cite{Wang2015f}}
\footnotetext[7]{Ref. \cite{Li2014h}}
\footnotetext[8]{Ref. \cite{Schneider2018a}}
\footnotetext[9]{Ref. \cite{Arora2016}}
\footnotetext[10]{Ref. \cite{Plechinger2016b}}
\footnotetext[11]{Ref. \cite{Zipfel2018a}}
\footnotetext[12]{Ref. \cite{Aivazian2014}}
\footnotetext[13]{Ref. \cite{Chen2019}}
\footnotetext[14]{Ref. \cite{Srivastava2015b}}
\footnotetext[15]{Ref. \cite{Liu2019a}}
\footnotetext[16]{Ref. \cite{Koperski2018}}
\footnotetext[17]{Ref. \cite{Koperski2015}}
\footnotetext[19]{Ref. \cite{Robert2017}}
\footnotetext[20]{Ref. \cite{Robert2020}}
\footnotetext[21]{Ref. \cite{Molas2019}}
\footnotetext[22]{Ref. \cite{Forste2020}}
\end{table*}

To apply this first principles approach, we first consider TMD monolayers since they are well-studied 
and therefore represent a good test case. However, previous attempts to calculate the g-factor of TMD monolayers without making assumptions about the orbital moment contributions were not very satisfactory \cite{Kormanyos2015,Wang2015f,Rybkovskiy2017} - a problem that the present approach can solve.
For the g-factors of A and B excitons Eqs.~(\ref{eqn:g_exp}), (\ref{eqn:gex}) and (\ref{eqn:gnk}) give $g^\mathrm{1L}_\mathrm{A,B} = 2 g_\mathrm{+K} = 2(\Delta \Sigma_\mathrm{+K} + \Delta L_\mathrm{+K})$, where $\Delta \Sigma_\mathrm{+K}$ and $\Delta L_\mathrm{+K}$ are the difference of the spin and the orbital angular momentum expectation values between conduction and valence band, respectively.
Figure \ref{fig:1L}(c) shows that circular polarized light couples valence and conduction band states with the same spin, consequently $\Delta \Sigma_\mathrm{+K} =0$ and only $\Delta L_\mathrm{+K}$ matters. 
In WS$_2$ the A (B) excitons are formed by the transitions v $ \rightarrow$ c+1 (v--1 $\rightarrow$ c) and therefore $g^\mathrm{1L}_\mathrm{A} = 2 (L_{c+1,+K} - L_{v,+K})$ and $g^\mathrm{1L}_\mathrm{B} = 2 (L_{c,+K} - L_{v-1,+K})$.  

Figure \ref{fig:conv}(a) and (b) show the convergence of $L_{n, \mathrm{+K}}$ and $g^\mathrm{1L}$ with respect to the number of bands $N$ included in the calculation (Eq.~(\ref{eqn:L})) for WS$_2$. 
The convergence behavior of the other considered TMD is shown in Fig.~\ref{fig:conv-all}.
The largest contribution to $L_{n, \mathrm{+K}}$ is at the band gap (dashed vertical line) because the energy denominator in Eq.~(\ref{eqn:L}) is smallest there, but apart from that, the convergence is very slow.  We find that for all considered TMD and the PBE-PAW method around $N = 300 - 500$ states are required to converge both quantities to a precision of 0.1 and around 700--900 to obtain an accuracy of 0.01 (for details see Fig.~\ref{fig:conv-all}).
The slow convergence can be understood by noticing that TMD monolayers strongly absorb light over a broad energy range \cite{Bernardi2013}, which means that there are many optical transitions with high intensities (momentum matrix elements) that contribute to Eq.~(\ref{eqn:L}).
This slow convergence is in contrast to conventional semiconductors, 
where only a few bands are required to obtain convergence \cite{Hermann1977a}.
This finally explains why previous attempts to calculate  exciton g-factors with few-band models did not lead to satisfactory results \cite{Kormanyos2015,Wang2015f,Rybkovskiy2017} -- the orbital contributions were not converged.

Figure \ref{fig:conv} also shows that for the same geometry the PBE and LDA results, obtained with the plane-wave-based, frozen-core PAW method (PBE-PAW and LDA-PAW) and the all-electron, full-potential  LAPW method (PBE-LAPW) are nearly identical. This shows that our results are consistent and not bound to a specific code or (semi)local functional; the small differences are due to numerical reasons. 

It is well-know that standard DFT calculations using (semi)local  functionals like PBE or LDA underestimate band gaps. This overestimates $L_{n, \mathrm{+K}}$, due to the energy denominator in Eq.~(\ref{eqn:L}). 
Quasiparticle GW calculations are able to correct this problem but they are numerically expensive. 
Fortunately the  wavefunctions obtained from (semi)local DFT are almost identical to GW wavefunctions \cite{delSole,Deslippe2012} (which explains why non-self-consistent approaches like G$_0$W$_0$ give reasonable results). Therefore we expect the DFT spin and momentum matrix elements $\Sigma_{nm \mathbf{k}}$ and ${\pi}^{\alpha}_{nm\mathbf{k}}$ to be reasonable and it is a good approximation to only correct the eigenvalue spectrum, in particular the band gaps. This is conveniently done by defining a "scissor operator"
\begin{align}
{\varepsilon^{0}_{n \mathbf{k}}}' = 
\left\{
\begin{array}{l}
  \varepsilon^0_{c \mathbf{k}} + \Delta \\
  \varepsilon^0_{v \mathbf{k}},
\end{array}
\right.
\end{align}
that modifies the band energies by simply increasing the band gap by $\Delta$.
{As} shown in Fig.~\ref{fig:conv}(c), $L_{n, \mathrm{+K}}$  decreases with $\Delta$. When increasing the band gaps of the considered TMD to their G$_0$W$_0$ value \cite{Rasmussen2015} (see dashed vertical line) the $L_{n, \mathrm{+K}}$ decrease by values ranging from 0.50--1.13 (17--44\%). These are big changes, which shows that calculating $L_{n, \mathrm{+K}}$ for individual bands is challenging. 
The individual g-factors of conduction or valence bands could be probed separately via transport experiments and this could provide some insight to identify the individual values.
However, the changes of the valence and conduction band states are very similar and when taking their difference for calculating the exciton g-factor, the band gap dependence nearly disappears. This is discernible in  Fig.~\ref{fig:conv}(d); $g^\mathrm{1L}_\mathrm{A,B}$ of TMD increase only by 0.15--0.18 (3.9--4.7\%) when the band gap is increased to the G$_0$W$_0$ value.
These changes are small enough 
to claim that standard DFT calculations using semilocal functional are suitable for calculating exciton g-factors. Therefore we do not apply the  "scissor operator" to the results below.

In Tab.~\ref{tab:1Ls} we provide the PBE-PAW g-factors  for the considered TMD, which are approximately equal to -4 for all systems. 
The experimental values, provided in the table, have a quite large statistical spread, even when we limit ourselves to undoped, encapsulated samples and measurements at $T=4$ K.
However, all values are negative and vary about -4, which is fully consistent with our theoretical results. 
To our knowledge, this represents the first successful, parameter-free calculation of exciton g-factors in TMD.
%
Overall, we do not find significant differences in the g-factors and the orbital angular momenta between the TMD monolayers. However, the calculated intensities in WX$_2$ are larger than in MoX${}_2$ systems, which is consistent with measured photoluminescence spectra at room temperature \cite{Koperski2017a}.
%
The orbital angular momenta at +K  in Tab.~\ref{tab:1Ls} are all positive and much bigger than commonly assumed in the literature, where $L$ is often approximated by the atomic orbital contribution ($L_{v, \mathrm{+K}} \approx L^\mathrm{ao}_{v, \mathrm{+K}} = 2$ and $L_{c, \mathrm{+K}} \approx L^\mathrm{ao}_{c, \mathrm{+K}}=0$) \cite{Srivastava2015b,Stier2016a,Koperski2017a}.
However, $\Delta L$ is always close to -2, which explains the success of these simple models.
%
In Tab.~\ref{tab:1Ls} the g-factors of both A and B excitons are given. The two values are quite similar and they are close to -4 in all systems. 
But we consistently find that $g^\mathrm{1L}_\mathrm{A} > g^\mathrm{1L}_\mathrm{B}$, which nicely agrees with experimental findings \cite{Stier2016a,Koperski2018}.

\subsection{Stacking- and spin-dependent g-factors of interlayer excitons in heterobilayers}

\begin{table*}
\caption{\label{tab:HBs} 
Calculated g-factors  $g^\mathrm{HB}$ (Eq. (\ref{eqn:g_exp})) of interlayer excitons for high-symmetry stacking configurations of MoSe$_2$/WSe$_2$ heterobilayers and comparison with reported experimental values. 
Also indicated are the corresponding transitions between the valence band (v) and the conduction (c, c+1) band at the +K point, their intensities $\frac{\hbar}{m_0} | \mathbf{e_\pm} \cdot \bm{\pi} |^2$ in (eV$\cdot${\AA})$^2$, 
circular polarizations and whether it is a spin-conserving ($\uparrow \uparrow$) or a spin-flip ($\uparrow \downarrow$) transition. 
$\Delta \Sigma = \Sigma_{c, \mathrm{+K}} - \Sigma_{v, \mathrm{+K}}$ is the spin contribution \Tom{(where $|\Sigma_{n,+\mathrm{K}}| = 1$ is used)}  and $\Delta L = L_{c,+\mathrm{K}} - L_{v,+\mathrm{K}}$ is the  orbital contribution to $g^\mathrm{HB}$; $L_n = L_{n,+\mathrm{K}}$. All results are obtained with the PBE-PAW method.
The g-factors are strongly stacking-dependent.  Good agreement with experiment is found for v $\rightarrow$ c transitions with sizable intensities (highlighted).
}
\begin{ruledtabular}
\begin{tabular}{lllllllll}
 & \multicolumn{2}{l}{$R^X_h$} & $R^h_h$ & $R^M_h$ & $H^X_h$ & \multicolumn{2}{l}{$H^h_h$} & $H^M_h$ \\
 \hline
$g^\mathrm{HB}$& \textbf{6.19}& -10.73& \textbf{-6.15}& 10.42& -12.60& \textbf{-16.67}& 12.15& 16.31 \\
$g^\mathrm{HB}$ (exp.) 
&6.72\footnotemark[1]& \Tom{-10.6}\footnotemark[6] 
&-8.5\footnotemark[2]&& 
&-15.89\footnotemark[1]& 10.7\footnotemark[4]\\ 
&7.1\footnotemark[2]&&&& 
&-15.1\footnotemark[3]\\   
&\Tom{6.99}\footnotemark[5] &&&&&-15.2\footnotemark[4] \\     
transition & v $\rightarrow$ c & v $\rightarrow$ c+1 & v $\rightarrow$ c & v $\rightarrow$ c+1 & v $\rightarrow$ c+1 & v $\rightarrow$ c & v $\rightarrow$ c+1 & v $\rightarrow$ c\\
intensity  &  0.08 &  0.05 &  0.12 & $10^{-7}$ & 0.01 & 0.03 & 0.34 & $10^{-4}$ \\
 polarization       &$\sigma-$ &$\sigma+$ &$\sigma+$ &$\sigma-$ &$\sigma+$ &$\sigma+$ &$\sigma-$ &$\sigma-$ \\

spin & $\uparrow \uparrow$ & $\uparrow \downarrow$ & $\uparrow \uparrow$ & $\uparrow \downarrow$ & $\uparrow  \uparrow$ & $\uparrow  \downarrow$ & $\uparrow  \uparrow$ & $\uparrow \downarrow$ \\
$\Delta \Sigma$  &  0     & -2    &  0     & -2    &  0     & -2    &  0     & -2    \\
$L_{c(+1)}$      & 1.80 & 1.53 & 1.79 & 1.53 & -1.53 & -1.79 & -1.53 & -1.78 \\  
$L_{v}$ & 4.90 & 4.90 & 4.86 & 4.74 & 4.77 & 4.54 & 4.54 & 4.37 \\
$\Delta L$ & -3.10  & -3.37 & -3.08 & -3.21 & -6.30 & -6.34 & -6.07 & -6.16\\
\end{tabular}
\end{ruledtabular}
\footnotetext[1]{Ref. \cite{Seyler2019}}
\footnotetext[2]{Ref. \cite{Ciarrocchi2019}}
\footnotetext[3]{Ref. \cite{Nagler2017}}
\footnotetext[4]{Ref. \cite{Wang2019}, the authors only measured $|g^\mathrm{HB}|$}
\footnotetext[5]{Ref. \cite{Joe2019}}
\footnotetext[6]{Ref. \cite{Joe2019}, value of charged exciton}
\end{table*}

Now we apply the method to interlayer excitons in van der Waals heterostructures. 
As prototypical moir{\'e} system we chose MoSe$_2$/WSe$_2$ HB  where unexpected values of g-factors were recently reported \cite{Nagler2017,Ciarrocchi2019,Seyler2019,Wang2019}. 
The lattice constants of the monolayers are almost identical and for precise twist angles of $\theta = 0^\circ$ (R) or 60$^\circ$ (H) (and multiples of it) the system is (quasi) commensurate \cite{Hsu2018}. But when samples are fabricated by exfoliation methods $\theta$  cannot be precisely controlled; for $\theta \approx 0^\circ$ or $\theta \approx 60^\circ$ the lattice reconstructs and certain high-symmetry stacking configurations dominate the sample (see Fig.~\ref{fig:HB}(a)) \cite{Rosenberger2019,Weston2019}. Thus it is sufficient to only study those high-symmetry stacking configurations, because they represent most of the properties of the HB. 

The calculated g-factors of K point interlayer excitons for each of these stackings are given in Tab.~\ref{tab:HBs}. These values show explicitly that g-factors in TMD HB are spin- and stacking-dependent, as discussed in Sec.~\ref{sec:TMD}. Also indicated are the corresponding optical transitions between the valence (v) and the conduction (c, c+1) bands and their intensities, which
are two to three orders of magnitude lower than the ones of monolayer transitions (see Tab.~\ref{tab:1Ls}). This agrees well with previous results \cite{Komsa2013a,Gillen2018,Yu2018} and explains why interlayer excitons are hard to observe by absorption spectroscopy and are typically probed in photoluminesce experiments.
The intensities of R$^M_h$ and H$^M_h$ are significantly lower and the transitions can probably not be observed. 
If we further consider that experimentally g-factors are determined by low-temperature photoluminescence spectroscopy where only the lowest energy transition (v $\rightarrow$ c) matters, we are left with interlayer exciton g-factors of +6.2, and $-6.2$ for  0$^\circ$ (R) and $-16.7$ for 60$^\circ$ (H) systems (highlighted in Tab.~\ref{tab:HBs}). 
Taking into account the large statistical spread of reported experimental g-factors (see Tab.~\ref{tab:1Ls}), these values are almost in quantitative agreement with recent experiments on MoSe$_2$/WSe$_2$ HB, where we are able to match our result to measurements of Seyler \textit{et al.} ($6.72 \pm 0.02$ for $\theta \approx 2^{\circ}$ and $-15.89 \pm 0.03$ for $\theta \approx 57^{\circ}$), Ciarrocchi \textit{et al.} ($+7.1 \pm 1.6$ and $-8.5 \pm 1.5$ for $|\theta| < 1^{\circ}$), Nagler \textit{et al.} ($-15.1 \pm 0.1$ for $\theta \approx 54^{\circ}$) \Tom{and Joe \textit{et al.} ($+6.99 \pm 0.35$) \cite{Seyler2019,Ciarrocchi2019,Nagler2017,Joe2019}}.
The g-factor of $-16.7$ originates from regions with H$_h^h$ stacking, which is also covering most of the sample (see Fig.~\ref{fig:HB}(a)). 
In R systems g=+6.2 is linked to R$_h^X$, which is the dominant stacking (together with R$_h^M$). The negative g-factor $-6.2$ comes from regions with R$_h^h$ stacking, that is present only in small parts of the samples (the nodes). 
Ciarrocchi \textit{et al.}~\cite{Ciarrocchi2019} ascribe their $g=-8.5$ peak to the spin-conserving and and the $g=+7.1$ peak to the spin-flip transition of the R$_h^X$ stacking. 
However, in Tab.~\ref{tab:HBs} the signs of the calculated g-factors of  spin-conserving and spin-flip transitions of R$_h^X$ are exactly opposite to their interpretation and the magnitudes of these two g-factors differ substantially.
Therefore our results suggest that the two peaks reported by Ciarrocchi \textit{et al.} are related to spin-conserving transitions and they originate from different parts of the sample.  
For H systems Wang \textit{et al.}~\cite{Wang2019} find two transitions with g-factor magnitudes of $|g^H|=15.2 \pm 0.2$ and $10.7 \pm 0.2$ (it is important to note that the authors did not determine the sign of their g-factors) and assign them to spin-singlet and spin-triplet excitons, which correspond to spin-conserving and spin-flip transitions, respectively. Our first principles results give slightly bigger magnitudes but otherwise confirm this assignment. The spin-conserving transitions of H$^X_h$ and H$^h_h$ are both  candidates to explain the lower of the two values; still it is more likely that the transition originates from H$^h_h$ because the samples are mostly covered by H$^h_h$ stackings and the oscillator strength of the transition is particularly large.
\Tom{In electron-doped R-type samples Joe \textit{et al.} measure a PL peak with $g^R=-10.6 \pm 1.0$ and in undoped samples they find $+6.99 \pm 0.35$ \cite{Joe2019}. The authors ascribe these two peaks to charged and neutral interlayer excitons, respectively. According to Wang \textit{et al.} the approach for calculating g-factors of neutral and charged excitons is the same \cite{Wang2015f}. Our values of $-10.7$ for the spin-flip (v $\rightarrow$ c+1) transition and $+6.2$ for the spin-conserving transition (v $\rightarrow$ c) transition in the R$^X_h$ stacking nicely agree with these measurements.  However more detailed analysis will
be necessary to fully understand the agreement for charged excitons.}
The remaining predicted values we present in Tab.~\ref{tab:HBs} could be observed in future experiments.

After showing the good agreement with recent experiments, let us now analyze orbital and spin contributions and the sign of the g-factors. 
In  MoSe$_2$/WSe$_2$ HB the band alignment is such that MoSe$_2$ states form the conduction band  and WSe$_2$ states the valence band. This is indicated by the color code in Figs.~\ref{fig:HB}(b,c). 
In TMD HB the K-point states do not hybridize and are basically a superposition of monolayer states \cite{Hanbicki2018}. That is why the magnitudes of $L_{n}$ in Tab.~\ref{tab:HBs} deviate only marginally from the corresponding monolayer values. In H systems the real space twist of the monolayers relative to each other  is connected to a similar twist of the Brillouin zones. Hence for H systems 
the MoSe$_2$ conduction band state from --K is at +K in the HB (see Fig.~\ref{fig:HB}(c)). This swaps the sign of the related spin and orbital contributions, as presented in Tab.~\ref{tab:HBs} by the negative value of $L_{c(+1)}$ for H systems. As a consequence the orbital contribution $\Delta L$ of H systems is approximately twice the value of R systems, which  explains why the magnitude of the g-factors is always bigger for H than for R systems.
In HB both spin-conserving ($\uparrow \uparrow$) and spin-flip transition ($\uparrow \downarrow$) can couple to circularly polarized light  
and hence they matter when defining the g-factor via Eq.~(\ref{eqn:g_exp}). Furthermore a spin-flip transition provides a spin contribution to the g-factor of $\Delta \Sigma = -2$, that generally increases the magnitude of the g-factor. 
This is most significant for the g-factor of -16.7 for the H$^h_h$ stacking configuration. The large magnitude is a consequence of it being (i) a H transition and (ii) a spin-flip transition (leading to a spin-triplet exciton) \cite{Wang2019}.
If we consider the \textit{intra}-valley g-factor at +K, as defined by Eq.~(\ref{eqn:gex}), all g-factors would be negative, because only $\Delta L$ and $\Delta \Sigma$ matter. However, the \textit{inter}-valley g-factor, according to Eq.~(\ref{eqn:g_exp}) and commonly used in experiment, employs valley selection rules for circularly polarized light.
The stacking- and spin-dependence of these selection rules is what leads to g-factors with both positive and negative signs. For example, the intra-valley g-factor at +K for the $R_h^X$ stacking is $g_\mathrm{+K} = \Delta L + \Delta \Sigma= -3.10$ and $g_\mathrm{-K} = +3.10$, due to time-reversal symmetry. Then applying the corresponding optical selection rules to obtain the inter-valley g-factor gives $g^{R_h^X} = g_{\sigma +} - g_{\sigma -} = g_\mathrm{-K} - g_\mathrm{+K} = +6.2$.  
In many HB samples multiple interlayer exciton peaks are experimentally found and not all of them can be explained by considering momentum direct K-point transitions. It is likely that momentum-indirect excitons are playing an important role in these systems \cite{Kunstmann2018}.

Let us now have a look at the electron g-factor. Jian \textit{et al.} reported a value of $+1.07 \pm 0.079$ at +K (and $-1.11 \pm 0.095$ at --K) but they were not able to determine if their sample is R or H  
\cite{Jiang2018a}. Using the results in Tab.~\ref{tab:HBs} and  Eq.~\ref{eqn:gnk} we obtain $g_{c,+K} = +2.8$ 
for R stackings and the same value with negative sign for H stackings. Considering that the orbital contribution is calculated without scissor correction, we expect the actual g-factor to be smaller. If we now assume that the sign convention of Jian \textit{et al.}~is consistent with ours, our results indicate that their system is of R type (i.e. $\theta \approx 0$). 
Thus g-factor measurements of excitons (or even electrons) combined with our results enable to determine whether a system is R or H. For exfoliated HB such a tool is sometimes needed, because the usual method of choice, i.e.~second harmonic generation measurements, is not always perfectly robust for such systems.

\section{Summary}
In this paper we showed that g-factors of excitons in semiconductors (value and sign) can be determined by first principles methods if the calculation of the orbital angular momentum $L$ is properly converged. 
For the considered two-dimensional materials  hundreds of bands were required to obtain reasonable convergence, indicating that the basis set size is a critical numerical issue. 
For an individual Bloch state the calculation of $L$ suffers from the well-known band gap underestimation of density functional theory. However, the error in $L$ is approximately the same for electron and hole states and for excitons (which depend on the  difference $\Delta L$) error cancellation enables quantitative calculations.

We applied the method to excitons in monolayers of semiconducting MX$_2$ (M=Mo, W; X=S, Se, Te) 
and interlayer excitons in MoSe$_2$/WSe$_2$ heterobilayers and obtain good agreement with available experimental data.  
The precision of our method allows to assign measured g-factors of optical peaks to specific transitions in the band structure and also to specific regions of the samples. This revealed the nature of various, previously measured interlayer exciton peaks. 
We further show that due to specific optical selection rules g-factors in van der Waals heterostructures are strongly stacking- and spin-dependent. 

The presented numerical approach can be applied to a wide variety of semiconductors. Combined with g-factor measurements it might become a useful tool that helps to reveal the nature of optical excitations in semiconductors.

\textit{Note added.} During the submission of this article two preprints on the calculation of exciton g-factors of TMD monolayers using first principles methods appeared \cite{Deilmann2020,Forste2020}.

\begin{acknowledgments}
T.W.~acknowledges financial support by the Polish Ministry of Science and Higher Education  via the "Diamond Grant" no. D\textbackslash 2015 002645. P.E.F.J.~acknowledges financial support from the Alexander von Humboldt Foundation, Capes (Grant No. 99999.000420/2016-06) and DFG SFB 1277 (Project-ID314695032). J.K. and G.S. acknowledge financial support by the German Research Foundation (DFG) under grant numbers SE 651/45-1. A.~C. acknowledges financial support from the Brazilian National Council of Research (CNPq), throught the PRONEX and PQ programs.
Computational resources for this project were provided by ZIH Dresden under project "transphemat". We thank Florian Arnold (TU Dresden) for help with illustrations and M.~M.~Glazov (Ioffe Institute, St.Petersburg, Russia), Tobias Korn (University of Rostock, Germany), Philipp Nagler, Johannes Holler, Jaroslav Fabian and Tobias Frank (University of Regensburg, Germany) for inspiring discussions.
\end{acknowledgments}

\appendix

\begin{figure}[tb]
\includegraphics[width=\columnwidth]{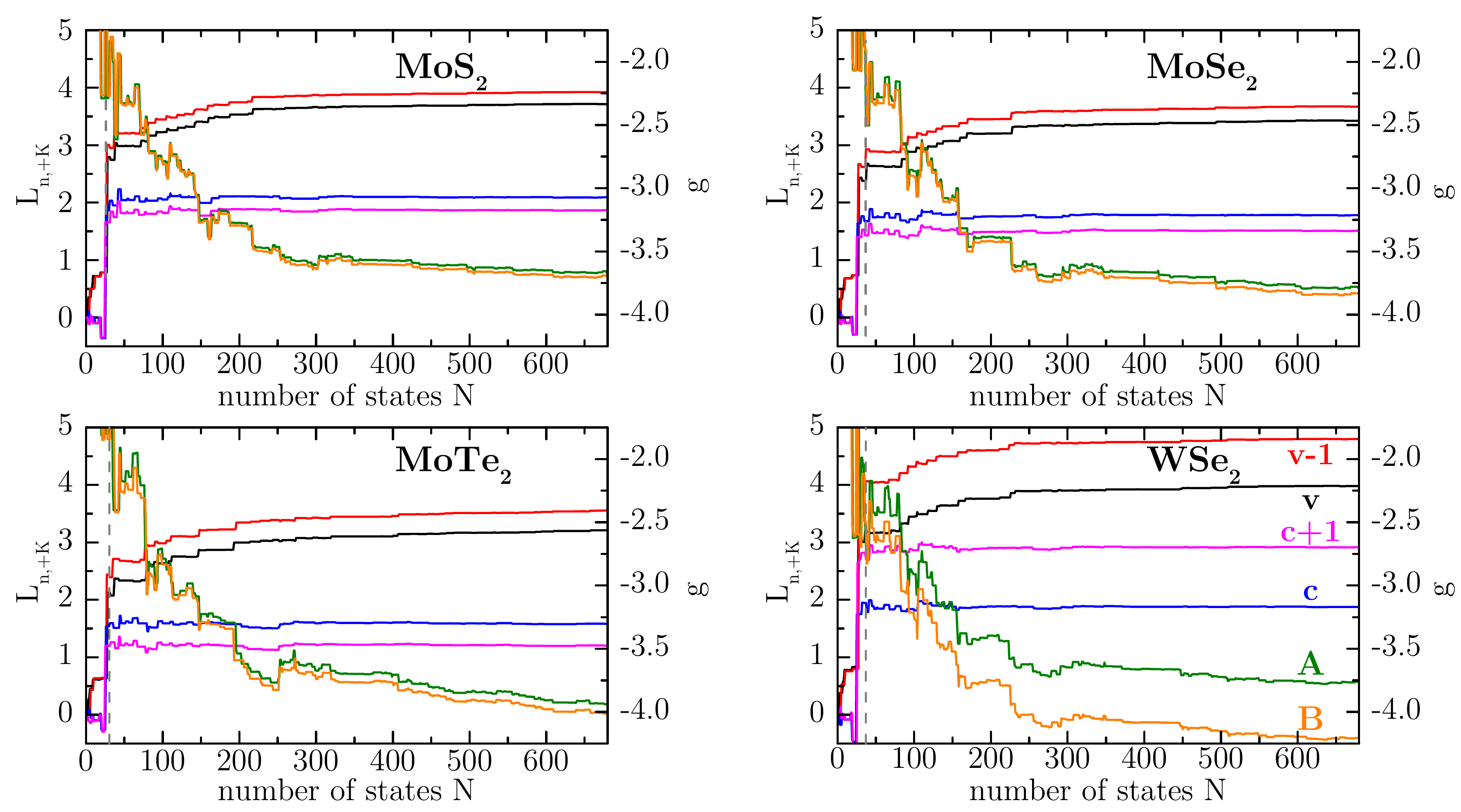}
\caption{\label{fig:conv-all} 
Convergence of the orbital angular momenta $L_{n,+\mathrm{K}}$ of the two highest valence band states ($n=v,v-1$) and two lowest conduction band states ($n=c,c+1$) at the +K point with respect to the number of states $N$ included in the calculation  and the convergence of the inter-valley exciton g-factor $g^\mathrm{1L} = 2 (L_{c(+1),+K} - L_{v(-1),+K})$ for A and B excitons in transition metal dichalcogenide monolayers. A large number of states $N$ is required converge the g-factors. For a precision of 0.1/0.01 $N=322/695$ states are necessary in MoS$_2$, 321/771 in MoSe$_2$, 547/881 in MoTe$_2$, 376/604 in WS$_2$ and 303/746 in WSe$_2$.
The g-factors of different TMD monolayers are similar and the value of the B exciton is always lower than that of the A exciton. In WSe$_2$ $g^\mathrm{1L}_\mathrm{A}$ and $g^\mathrm{1L}_\mathrm{B}$ differ the most (see Tab.~\ref{tab:1Ls}). $N=1$ is the lowest-energy state of the valence shell, the \Tom{valence band maximum} is indicated by a dashed vertical line and all values are PBE-PAW results. 
}
\end{figure}



\bibliography{references}

\end{document}